# Modelling and Control of Subsonic Missile for Air-to-Air Interception

**Rory Jenkins, Xinhua Wang**
Aerospace Engineering,
University of Nottingham, UK
Email: wangxinhua04@gmail.com

**ABSTRACT**

Subsonic missiles play an important role in modern air-to-air combat scenarios - utilised by the F-35 Lightning II - but require complex Guidance, Navigation and Control systems to manoeuvre with 30G's of acceleration to intercept successfully. Challenges with mathematically modelling and controlling such a dynamic system must be addressed, high frequency noise rejected, and actuator delay compensated for. This paper aims to investigate the control systems necessary for interception. It also proposes a subsonic design utilising literature and prior research, suggests aerodynamic derivatives, and analyses a designed 2D reduced pitch autopilot control system response against performances. The pitch autopilot model contains an optimised PID controller, 2$^{nd}$ order actuator, lead compensator and Kalman Filter, that rejects time varying disturbances and high frequency noise expected during flight. Simulation results confirm the effectiveness of the proposed method through reduction in rise time (21%), settle time (10%), and highlighted its high frequency deficiency with respect to the compensator integration. The actuator delay of 100ms has been negated by the augmented compensator autopilot controller so that it exceeds system performance requirements (1) & (3). However, (2) is not satisfied as 370% overshoot exists. This research confirms the importance of a lead compensator in missile GNC systems and furthers control design application through a specific configuration. Future research should build upon methods and models presented to construct and test an interception scenario.

## NOMENCLATURE

| | |
|---|---|
| $a_{xb}$ | x-component Acceleration wrt Missile body |
| $a_{yb}$ | y-component Acceleration wrt Missile body |
| AR | Aspect Ratio |
| $A_e$ | Nozzle exit area |
| b | Wingspan |
| $C_L$ | Coefficient of Lift |
| $C_{D0}$ | Base Drag Coefficient |
| $C_{L\alpha}$ | Lift Curve Coefficient |
| $C_{M\alpha}$ | Pitching Moment Coefficient |
| $C_{L\delta}$ | Elevator Lift Coefficient |
| $C_{M\delta}$ | Elevator Moment Coefficient |
| $C_{MAC}$ | Length of Mean Aerodynamic Chord |
| D | Drag |
| d | Diameter |
| G | G-Force acceleration ~ 9.81ms$^{-2}$ |
| h | Operating Altitude |
| L | Lift |
| $l_M$ | Missile Length |
| $l_N$ | Nose Length |
| $l_B$ | Body Length |
| $l_{BT}$ | Boattail Length |
| $J_Z$ | Moment of Inertia |
| Ma | Mach Number |
| m | Missile Mass |
| P | Thrust |
| $S_{ref}$ | Reference Surface Area |
| $S_W$ | Wing Surface Area |
| $S_T$ | Tail Surface Area |
| $V_c$ | Velocity at cruise |
| $\omega$ | Missile Pitch angle |
| $X_{AC}$ | Distance to Aerodynamic centre |
| $X_{MAC}$ | Distance to Mean Aerodynamic chord |
| $X_{CG}$ | Distance to Centre of Gravity |

## INTRODUCTION

This paper details the design of a subsonic missile for air-to-air interception scenarios. The overall missile configuration is proposed using existing designs, established literature by E. L Fleeman [1], and research by N. Sugendran [2] for elevator sizing. The design was then modelled in Computer Aided Design software (CAD). Applicable aerodynamic derivatives are estimated and/or obtained from research papers by E. L. Fleeman [1], X. Wang et al [3], as well as J. O. Nichols [4] for use in the control models. However, the results are not directly applicable to the specific Mach design point in question and limited to low angles of attack.

The control systems for a reduced 2D Missile body dynamic description was derived, using modern literature by L. Defu et al [6], reducing the complexity of problem after considering E. L. Duke et al [5] from 3D to 2D.

The Pitch autopilot design utilised research by Carey et al [9], K. Nirmal et al [10], Shima et al [11], Wang et al [12]. Initial design of the actuator utilised parameters from T. Harold [8], however this was later revised. The control systems of a missile are crucial, and contain Guidance, Navigation and Control methods (GNC) which include Guidance Laws, Hinge moment autopilot, Pitch dynamic PID control, and translational dynamic controllers.

The pitch autopilot was augmented with a lead compensator to improve the system response. The results with and without compensator are discussed and evaluated against performances established by E. Devaud [14].

Research into the Hinge Moment Autopilot was conducted. Future work should continue to build upon the proposed description, and utilise blending techniques researched by R. A. Hyde [14] to combine all systems to exceed a normal acceleration of 30G's as suggested by J. A. Kaplan [7] to ensure interception.

## PAPER FORMAT

This paper is subdivided into 4 sections which explore the breadth of design, shown in red in Figure 1.

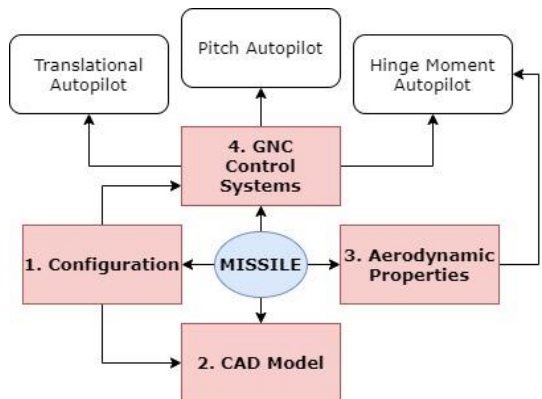

**Figure 1: Dissertation Structure**

The configuration properties of inertia, elevator geometrics, aerodynamic derivatives and dynamic pressure are utilised by the hinge moment autopilot. The pitch autopilot requires the missile moment of inertia. The translational autopilots require drag and lift estimations.

## 1. MISSILE CONFIGURATION

The configuration of the proposed subsonic missile is explored in this subsection. An initial understanding was developed by looking at existing configurations, such as the Raytheon AIM-9X-2 & MBDA ASRAAM AIM-132 which are actively used in Air-to-air combat situations. Tactical missile design by E. L. Fleeman et al [1] was used in configuring the missile for the Mach number design point. The proposed design was then constructed in 3-Dimensional digital space using CAD software by Dassault Systèmes for visualisation.

### 1.1 Existing Solutions

Missiles can be classed by velocity, which is broken down into subsonic, supersonic, and hypersonic. Subsonic missiles are relatively cheap per unit (~£400,000) are small, have relatively low velocities but are highly manoeuvrable and can destroy aircraft costing lives and tens of millions of pounds. The missile proposed in this paper is typical of ones carried by most fighter aircraft (e.g., the modern 2020 F-35 Lightning II which carries two AIM-9X missiles). Configuration data for two UK missiles are shown in Table 1 and were used initially in sizing the design configuration.

**Table 1: UK Existing Air-to-Air Missiles**

| Missile name | Weight [kg] | Range [km] | $l_M$ [mm] | d [mm] | Wings pan [mm] |
|---|---|---|---|---|---|
| Raytheon AIM-9X-2 | 85 | 18 | 3000 | 127 | 353 |
| MBDA ASRAAM AIM-132 | 88 | 50 | 2900 | 166 | 450 |

### 1.2 Missile Configuration

The proposed subsonic missile is winged, tail-controlled missile with an axisymmetric configuration. The design configuration areas explored relate to the body, nose, boattail, main wing and elevators. Missiles operating in transonic region and supersonic speeds (0.8<Ma<1.2) experience a substantial increase in the total drag on aircraft due to fundamental changes in the pressure distribution, highlighted in Figure 2. Configurations that favour drag reduction and that prioritized aerodynamic efficiency were selected for the design point of Ma = 0.85.

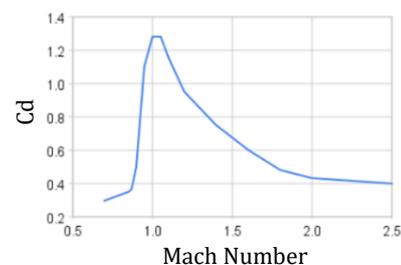

Mach Number

**Figure 2: Variation of Cd with Mach**

<u>Body Design</u>
This missile is circular (cross-sectionally) a/b = 1, using the wing to provide lift, as opposed to the body being a lifting surface (a/b>1). Future designs could improve on this to test if a lifting body surface is better suited to this Mach number scenario, versus a design utilising wing lift only. A body fineness ratio for subsonic class typically falls within 5< l/d < 25, with this missile having $l_M/d$ = 20. A high fineness ratio is desirable to reduce skin friction drag. With a diameter of 200mm, a body length of 4000mm was determined.

### Nose Design

A 'tangent ogive' nose design was selected due to a high nose fineness ratio as it is aerodynamically superior, with a value of $l_N/d=5$.

### Boattail Design

When motor burn occurs during powered flight, the base area that is outside the nozzle has a pressure larger than that of the free stream. A boat-tail design can reduce base drag by around 50% [1]. A selected design slope of 9° prevents flow separation, with $d_{BT}/d_{ref} = 1$ & $l_{BT}=200$mm for a spherical rear. It has a nozzle exit area of 0.015m$^2$.

### Wing Configuration

Having a large wing has advantages of range in subsonic flight, range in low dynamic pressure, lower guidance time constant, higher normal acceleration, high-altitude intercept, body stiffness and seeker tracking, all applicable to the design point. Furthermore, 70% of subsonic missiles have high aspect ratio wings [1]. A trapezoidal wing was deemed most suitable, due to stability and control. When sizing the wing, equation (1) [1] was used which utilizes Slender Wing Theory and Newtonian Impact Theory, suitable at subsonic flight. An aspect ratio of AR = 2.75 was selected which accounts for the overprediction by slender wing theory Using Equation 1, for a fin diameter of 444mm, the Wing surface Area ($S_W$) was 0.287m$^2$.

$$AR = b^2/S_W \qquad (1)$$

At subsonic flight the longitudinal centre of pressure ($X_{AC}$) is assumed to be located at 25% of the $C_{MAC}$. The Xac of the missile is located at 53% along the body, (63% from the nose length is typical for a missile with no flare at low angles of attack) [1], so configuration 10% away from the norm was acceptable. The missile has Lac = 3150mm, which satisfies the static margin of 10%.

### Static Stability

Main wing control was deemed unsuitable, and so tail elevator control was selected. A statically stable missile was designed, which has the aerodynamic centre ($X_{AC}$) located aft of the centre of gravity ($X_{CG}$). This means a nose up (increase in angle of attack) causes a negative pitching moment (nose down), which can then be trimmed by the elevator. This elevator induces a nose up manoeuvre and is controlled through the hinge moment autopilot. It can converge to the desired angle of attack (AOA) and designed so that it converges optimally. An unstable missile has the opposite characteristics and requires complex control systems to control the missile, else large oscillations occur, and/or divergence. A statically unstable missile has high responsiveness and manoeuvrability but requires a large elevator to provide moments, which increases drag. A static margin of 5-15% is desirable, so 10% was selected.

The $X_{CG}$ is located at 50% of the length of the missile from the nose tip, which is at 2500mm. The tail surface was then sized to provide static stability.

### Tail Elevator Sizing

Tail stabilizers have lower drag advantages than flare stabilizer design missiles [4]. The tail contribution to pitching moment must balance the pitching moments from the nose and wing, so Equation (2) was used to size the required tail area for a given static margin [1, 2]. The missile design angle of attack was selected at 0°. All 'X' distances were taken relative to the nose of the missile.

$$\frac{S_T}{S_{Ref}} = (C_{N\alpha})_{Body} \left\{ \frac{[X_{CG}-(X_{CP})_{Body}]}{d} \right\} + (C_{N\alpha})_{Wing} \left\{ \frac{[X_{CG}-(X_{CP})_{Wing}]}{d} \right\} \left( \frac{S_W}{S_{Ref}} \right) + \left\{ \left[ (C_{N\alpha})_{Body} + (C_{N\alpha})_{Wing} \left( \frac{S_W}{S_{Ref}} \right) \right] \left[ \frac{X_{AC}-X_{CG}}{d} \right] \right\} / \left\{ (C_{N\alpha})_{Tail} \frac{[(X_{AC})_{Tail} - X_{CG}]}{d} - \frac{(X_{AC}-X_{CG})}{d} \right\} \qquad (2)$$

Assumptions and reductions made when calculating Equation (2) are tabulated in Table 2, with respect to prior research by N. Sugendran [2].

**Table 2: Tail Sizing Assumptions**

| d | Diameter | 200 mm |
|---|---|---|
| $S_W$ | Wing Surface area | 0.282 m$^2$ |
| $X_{CG}$ | $L_M/2$ | 2600 mm |
| $S_{Ref}$ | $\frac{\pi}{4}d^2$ | 0.0314 m$^2$ |
| $(X_{CP})_{Body}$ | Slender wing theory: ~ d | 200 mm |
| $X_{CG} - (X_{CP})_{Wing}$ | 10% static margin | -250 mm |
| $(X_{CP})_{Wing}$ | ~50% of $C_{MAC}$ | 188.5 mm |
| $(X_{CP})_{Tail}$ | ~ l - d | 4800 mm |
| $(C_{N\alpha})_{Body}$ | 2 per radian of AOA | 0 (@ 0°AOA) |
| $(C_{N\alpha})_{Wing}$ | Slender wing theory: $\frac{\pi}{2}A_W$ | 0.262 |
| $(C_{N\alpha})_{Tail}$ | Slender wing theory: $\frac{\pi}{2}A_T$ | 0.262 |

Therefore, $\frac{S_T}{S_{Ref}}$ = -2.754 (3d.p), meaning a Tail surface area of 0.0865 m$^2$ was required.

This 2.75 value is large when compared to Figure 2.48, p75 [1], who suggest that it should lie within region of 0.8-1.5 for a Ma = 1.

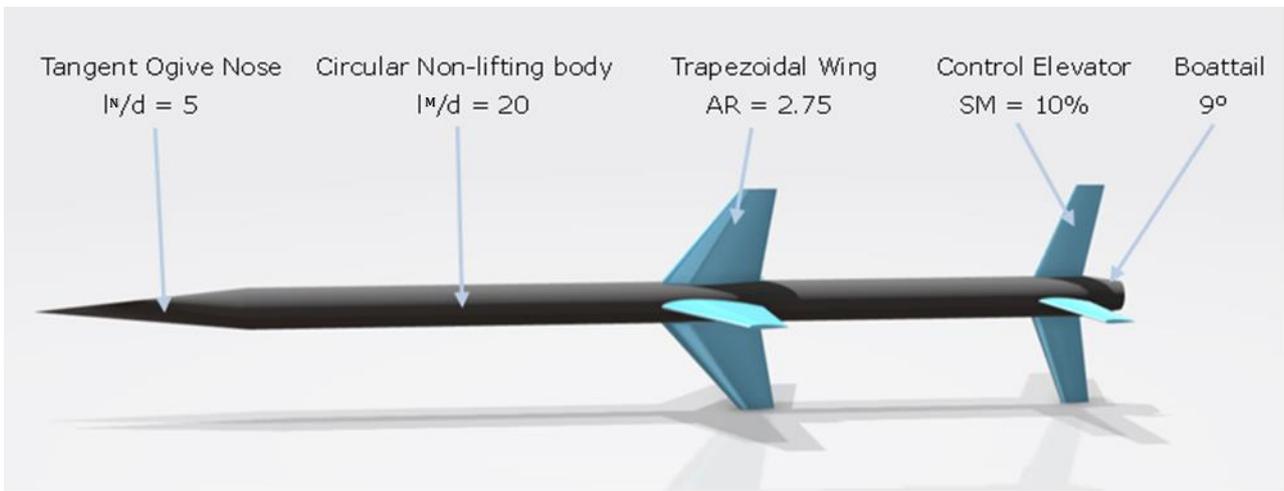

**Figure 3: Proposed Subsonic Missile CAD Render**

### 1.3 Configuration Design Summary

The proposed missile has the following configuration parameters, summarised in table 3 and labelled in Figure 3.

**Table 3: Missile Configuration Summary**

| Missile Parameter | Value | Unit |
|---|---|---|
| AR | 2.75 | - |
| $A_e$ | 0.015 | m² |
| b | 444 | mm |
| $C_{MAC}$ | 377 | mm |
| d | 200 | mm |
| h | 6000 | m |
| $J_z$ | 40 | kg/m² |
| $L_M$ | 5200 | mm |
| $L_{AC}$ | 3150 | mm |
| $l_N$ | 1000 | mm |
| $l_B$ | 4000 | mm |
| $l_{BT}$ | 200 | mm |
| $V_c$ | 250 | m/s |
| Ma | 0.85 | - |
| m | 85 | kg |
| $S_W$ | 0.282 | m² |
| $S_T$ | 0.0865 | m² |
| $X_{AC}$ | 94 | mm |
| $X_{MAC}$ | 2750 | mm |
| $X_{CG}$ | 2500 | mm |

## 2. CAD MODEL

The proposed missile was constructed in digital 3-Dimensional space using Catia 3D Experience CAD software by Dassault Systèmes - utilised within the modern Aerospace industry - exhibited in Figure 3. The design used the following apps in the software: Assembly Design, Part Design, Generative Wireframe & Surface, Live Rendering.

A top-down design approach was taken, with the missile design tree featuring a major assembly, constraints, then broken down into sub-assemblies and parts, which contained the major design features.

E.g., Subsonic Missile Assembly > Missile Main Wing Assembly > Aerofoil Original Part, References Set, Connections etc.

A master 'references nose' set was created first which contained all geometry points with respect to the nose tip, with which design sketches referenced. Creating a 'references' geometry set within each sub-assembly that contained critical planes, points and other features meant changes could be made quickly and easily. The wing and elevator were mirrored to create the four surfaces using a crown pattern tool as the missile is axisymmetric about the longitudinal axis. A scaling factor was used to determine lengths for the rear elevator, by resizing the main wing geometry. A thin NACA aerofoil was imported as a placeholder.

## 3. AERODYNAMIC DERIVATIVES

Aerodynamic derivatives for this missile are required by the translational and hinge moment autopilots. Aerodynamic lift and drag data for this paper was estimated using slender wing theory & Newtonian impact theory applicable for low subsonic flight [1]. For the designed missile configuration. These estimates are only applicable for low angles of attack ($\alpha < 10°$). The data in Table 4 is shown for an $\alpha = 0°$.

Applicable Elevator derivatives from a paper by Wang et al, in which an agile tail-sitter aircraft was developed were utilised [3], as well as aerodynamic data taken from J. O. Nichols [4] in which supersonic air-to-air missile data was compiled. Although Ma=2.3 is not applicable to this design scenario, the nature of defence classifications meant there was a scarcity of viewable sources. The $C_{M\alpha}$ derivative must be negative for a statically stable missile, so that a restoring moment will be generated to help reduce the angle of attack and stabilize the missile body [4].

**Table 4: Proposed Missile Aerodynamic Derivatives**

| Aerodynamic Derivative | Value | Reference |
|---|---|---|
| $C_L$ | 0.924 | [1] |
| $C_{D0}$ | 0.603 | [1] |
| $C_{L\alpha}$ | $0.524 + 2(\alpha)$ | [1] |
| $C_{M\alpha}$ | -0.300 | [3] |
| $C_{L\delta}$ | 0.208 | [4] |
| $C_{M\delta}$ | 0.267 | [4] |

E. L. Fleeman states that $C_{M\alpha} < C_{M\delta}$ pitching moment from angle of attack should be less than the elevator control effectiveness to have adequate control margin [1], this is satisfied.

## 4. GNC CONTROL SYSTEM DESIGN

In this section the missile's mathematical model for body dynamics are derived, pitch autopilot presented and analysed with respect to performances. The missile autopilot must reject expected disturbances. Simulink within Matlab 2021a software by MathWorks was used to model and analyse expected performances of the designed systems.

The block diagram shown in Figure 4 details the 2D missile GNC control system, showing the relationship of pitch dynamic autopilot, which is contained within the hinge moment autopilot. The Missile longitudinal dynamic system contains the missile acceleration $a_{xb}$ parameter to be controlled. These systems generate body accelerations that are interpreted by the Navigation and Seeker into positional information, which is input to update the Guidance Laws.

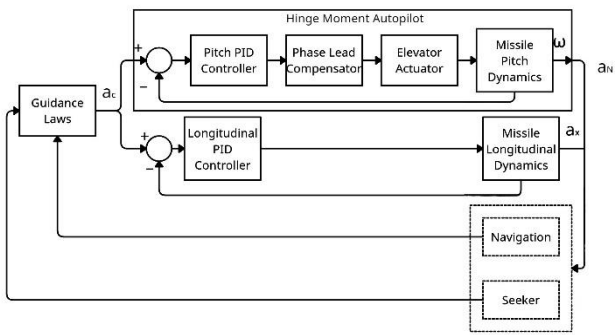

**Figure 4: Missile GNC Loop Block Diagram**

### 4.1 Missile Plant Dynamics

The missile dynamic model equations needed to be defined and selected so that they could be controlled. Research by NASA (E. L. Duke et al) [5] showed the full extent of 3D coupled non-linear aircraft dynamics requiring to be solved. Literature by L. Defu et al [6] was used to help reduce the complexity from a 3-Dimensional (3D) aircraft motion description described by six nonlinear simultaneous equations (6 DOF needing to be solved for) to a 2D coupled dynamic model with 3 DOF to be solved. Figure 5 depicts the missile overview. Each DOF equation will require linearising and a PID loop to regulate. Table 4 tabulates nomenclature applicable to the proposed missile, and the reduction from 3D to 2D.

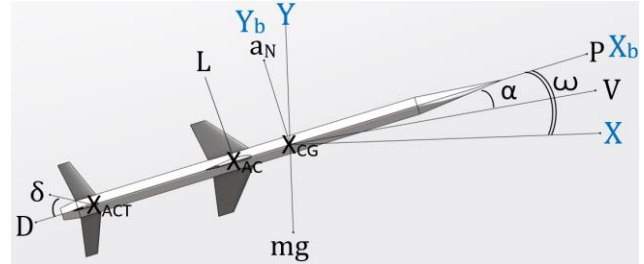

**Figure 5: Subsonic Missile**

**Table 5: 2D Reduced Coordinate References**

| Missile body coordinate system | Roll | Yaw | Pitch |
|---|---|---|---|
| Reference Axis | $X_b$ | $Y_b$ | $Z_b$ |
| Angular pitch | / | / | $\omega$ |
| Velocity component | $V_{xb}$ | $V_{yb}$ | / |
| Forces acting on the missile | $X_b$ | $Y_b$ | / |
| Moment acting on the missile | / | / | $M_z$ |
| Moment of inertia | / | / | $J_z$ |
| Product of inertia | / | / | / |

To simulate a scenario, properties of the missile had to be finalised. For the reduced order model, the following assumptions were made:
- Mass is fixed at 85kg during operation
- Jz value is assumed to be constant
- Products of inertia omitted for axisymmetric model in 2D
- Main wing generates all lift for flight
- Rear elevator deflection creates torque to influence missile pitch angle

The motion of the missile is represented by three linearised simultaneous equations, 2 translational and 1 rotational that can then be controlled through PID techniques and stabilised. The PID controller reduces the error of the system to zero with minimum oscillations so that the missile can intercept its target missile successfully, stipulated by (J. A. Kaplan et al) [7] to be effective through shrapnel within ±10m.

For a missile to intercept, it is essential for the missile to constantly acquire motion information of the target in flight and adopt a target tactic. A review of Design Guidance and Control Systems for Tactical Missiles by L. Defu et al [6] gave an introduction to the guidance loops and techniques employed in modern systems. Roll autopilot and Yaw autopilot are not necessary, as the missile is confined to manoeuvres in the X and Y direction only, so subsequent lateral controllers were omitted.

<u>Force Balance Equations</u>

When a 3 DOF model of the missile is given in the body coordinate frame, the derived translational equations are reduced to (3) and (4). With the missile having a resultant vector of (5).

$$a_{xb} = \frac{1}{g*m}[P - D\cos\alpha + L\sin\alpha] \quad (3)$$

$$a_{yb} = \frac{1}{g*m}[D\sin\alpha + L\cos\alpha] = a_N \quad (4)$$

$$V = \sqrt{V_{xb}^2 + V_{yb}^2} \quad (5)$$

<u>Rotational Equation of Motion</u>

$$\dot{\omega} = \frac{a_N}{V} \quad (6)$$

The equation (6) contains omega dot, which is the pitch angular rate described in the missile rotational equation of motion. '$a_N$' is the normal acceleration, described by equation (4). The normal acceleration leads to a change in the missile's velocity vector and its flight path. Forces that act on the missile cause it to pivot about its centre of mass and are countered by the elevator deflection to adjust trajectory. These forces give rise to $V_{xb}$ and $V_{yb}$ components. Since the elevator is manoeuvred by the GNC system to change trajectory, this influences lift and drag, affecting the acceleration in the translational equations of motion. Research by J. A. Kaplan et al stipulate that current subsonic missiles maximise up to 30G's of normal acceleration when manoeuvring for interception [7]. This is only achievable once a sufficient velocity is reached following a rapid burn, which reduces the missiles weight significantly.

## 4.2 Actuator Modelling

The missile actuator converts the desired control command developed by the autopilot into rotational motion. Modelling the response of the rear elevator actuator gave a better description of the actual interception system. Initially the designed actuator parameters, were derived from [8] utilising a DC motor vibration suppression model parameters, which consumes low power. However, iterations and improvements explored in Section 4.4 meant the actuator parameters were later reassessed.

Electrical actuators are common today and actuate the control surfaces to give a desired response for the controller, whilst minimising packing space (T. Harold et al) [8]. The actuator is also the main component that limits the bandwidth of the missile autopilot [6]. A 2nd order dynamic model was selected, with the transfer function of the actuator described by equation (7) [6].

$$\frac{\delta(s)}{\delta_c(s)} = \frac{n*\omega_n^2}{s^2 + 2\mu\omega_n s + \omega_n^2} * e^{-\tau s} \quad (7)$$

Where $\delta_c$ is the actuator command supplied by the controller, and $\delta$ is the actual deflection provided by the actuator model. $\omega_n$ is the undamped natural frequency, $\mu$ the damping coefficient and $\tau$ the time delay of the actuator response, 'n' is the gain. The actuator amplifies the controller input through external energy - gain value - to create powerful torque.

## 4.3 Pitch Autopilot Control

Control research derived by (Carey et al) [9] was used to help create an autonomous control system for the pitch dynamics. The PID controller overcomes the time-varying disturbance, which simulates variance in drag due to wind and constant wind resistance. The purpose of this negative feedback loop autopilot is to tend the system error to zero, by comparing the current angle to the desired input.

The control plant contains the rearranged rotational equation (8).

$$\ddot{\omega}(t) = \frac{1}{J_z}c(t) - \frac{\lambda}{J_z}\dot{\omega} - \frac{d(t)}{J_z} \quad (8)$$

Where d(t) is the time varying disturbance, and $\lambda$ is the constant disturbance of air resistance. The control torque generated c(t) is operated by the aft horizontal elevator control surface, which is controlled by the pitch controller. Ultimately the hinge moment autopilot loop controls the desired pitch angle, in order to generate $a_N$ (6) to manoeuvre for interception, supplied with acceleration commands from the Guidance Laws. The disturbances. The time varying disturbance was simulated through a sine wave block with an amplitude of 1 and frequency of 1 rad/s.

The negative feedback control system derived contains the PID controller, lead compensator, elevator actuator, pitch dynamics and a Kalman filter. Any control system should also be able to reject HF noise, introduced through outputs from the Inertial Measurement Unit (IMU). This measures the three body inertial attitude angles as well as accelerations. For this 2D reduced system, IMU outputs pitch angular rate ($\dot{\omega}$) and body accelerations ($a_x$, $a_y$). (K. Nirmal et al) [10] characterise these random stochastic errors from the IMU arising due to measurement noise, drifting biases, and turn-on to turn-on bias variance. The HF was simulated via a Random Noise block, which generates a stochastic signal with a mean of 0, variance was set to 0.1, with a sample time of 0.01.

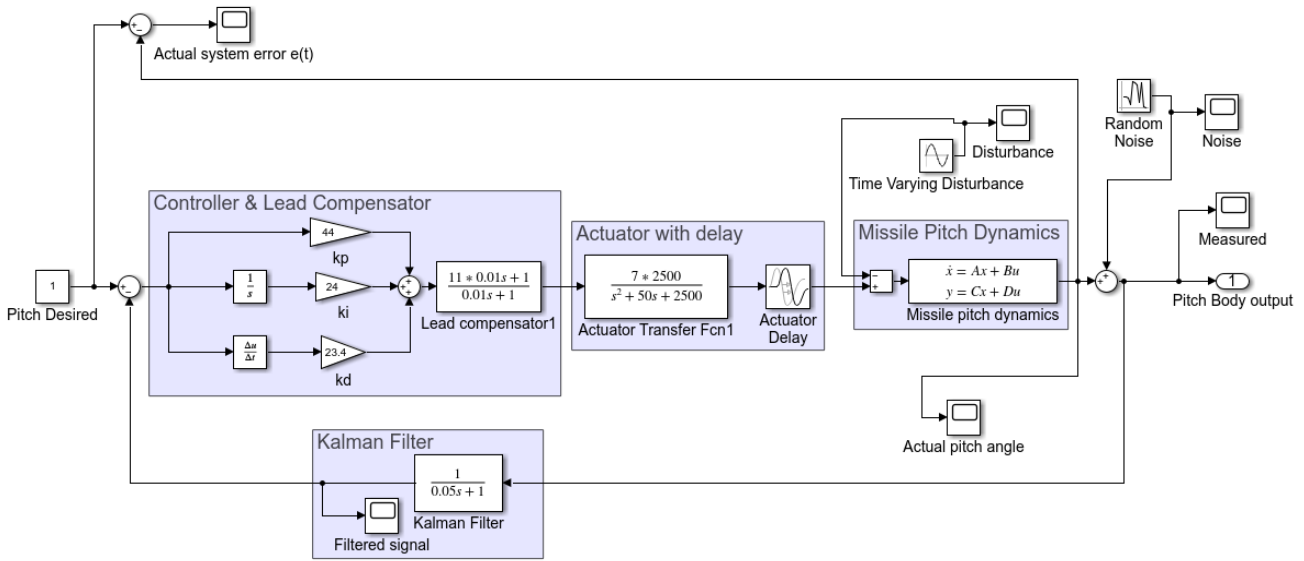

**Figure 6: Missile Pitch Autopilot 'B'**

Simulation tests were conducted from a starting Pitch angle of 10° to a desired angle of 1°. The simulation timing was confined to 10s maximum. Accurate measurements of the signal response times were carried out using the 'cursor measurement' tool built into Simulink.

An initial simulation tested the implementation of the Actuator and delay on the simple PID system. The plot of the initial system response is shown in Figure 7, which is stable. High frequency noise was not introduced in this initial test. Then the compensator was added to augment ability, and HF noise introduced. This final system was tested, with results detailed in 4.5.

A summary of the pitch autopilot parameters is shown in Table 6 for the final constructed pitch autopilot control system in Simulink - shown in Figure 6. The PID controller parameters were manually tuned to provide an optimum response by the system, so that the pitch tended to the desired angle with minimum oscillations.

**Table 6: Pitch Autopilot Parameters**

| Parameter | Nomenclature | Value |
|---|---|---|
| Proportional Gain | $k_P$ | 44 |
| Derivative Gain | $k_D$ | 24 |
| Integral Gain | $k_I$ | 23.4 |
| Aerodynamic Resistance | $\lambda$ | 6 |
| Time Varying Disturbance | $d(t)$ | 1 |
| Moment of Inertia | $J_z$ | 40 |

(Shima et al) [11] introduce Missile time constants for a canard actuated servo system $\tau_S$ = 20ms, missile dynamics as $\tau_M$ = 100ms, with target dynamics as $\tau_T$ = 50ms. A suitable delay of $\tau$ = 100ms was finalised for the autopilot, as targeting was not included within this system.

Initial testing of the pitch autopilot system showed that the system was unstable unless $\tau$<50ms was satisfied. This is not suitable, and the system required improvement.

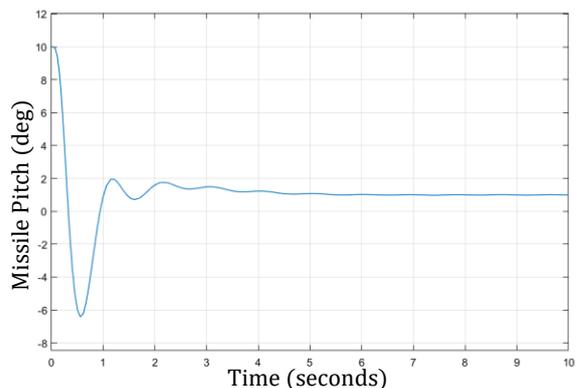

**Figure 7: Initial System Pitch Response**

## 4.4 Phase Lead Compensator

A general phase-lead compensator can be used as a signal prediction method to estimate the future signal f(t+Δ) and its derivatives (Wang, 2020) [12]. For successful missile interception the future prediction is necessary to increase the pitch ability of the defence missile and reduce manoeuvre time by compensating for the actuator delay response. The attack speed is significant, Vc = 250m/s, and so the usual methods of real-time interception control require augmentation. (L. Defu et al) [6] state that a lead compensator can be introduced to improve the actuator mechanism stability. Consequently, the settle time and oscillations are minimised, and the results verify this.

Subsequently the pitch angle autopilot controller was modified with the lead compensator. Limitations with the previous actuator design meant that a delay of 200ms created instability, even with the compensator present. In order to reduce oscillations, the model of actuator was selected as shown in equation (9), with a gain of 7 and a delay of $\tau = 100$ms. The gain was varied between 1 to 15, with 7 producing an optimal response.

$$\delta_0(s) = \frac{\delta(s)}{\delta_c(s)} = \frac{7*2500}{s^2 + 50s + 2500} * e^{-0.1s} \quad (9)$$

The phase lead compensator was implemented, described by equation (10) [12].

$$\frac{Y(s)}{F(s)} = \frac{aTs+1}{Ts+1} = \frac{11 \times 0.01s+1}{0.01s+1} \quad (10)$$

## 4.5 Results Discussion

For the proposed pitch control system in Figure 5, simulation results of the actual pitch response from $\omega = 10°$ to 1° is shown for the system with and without the compensator. The simulation results are presented through Figures 9 - 15. With Figures 9, 11, 13 for the system response without the compensator 'A', and Figures 8, 10, 12, 14 for the system response with compensator 'B'. Error tolerance lines for the steady state system response are labelled as horizontal lines 'Upper 1' & 'Lower 2', to visualise requirement (3) of 5%.

Figures 14 and 15 show the actual pitch error with plotted tolerance lines of ±0.45° (5%), with response settle time '$t_s$' reached when first entering this limit and remaining within the tolerances. Requirements (1) & (3) are satisfied for the pitch autopilot proposed, as presented by E. Devaud [13]. The actuator delay has been negated by the integration of the compensator into the system.

The actuator delay can be seen, where all responses remain at 10° for 100ms. Table 7 results show the ability of the phase-lead compensator to improve the stability and response of the proposed system model. It is a significant improvement compared to the initial result in Figure 7, where $M_p$ = 8.2°, $t_s$ =3600ms was achievable only with an unrealistically small actuator delay ($\tau$<50ms).

**Table 7: Pitch Autopilot Simulation Measurements**

| Measurement | Note | A | B |
|---|---|---|---|
| Percentage Overshoot | $\%M_P$ | 790% | 370% |
| Peak Overshoot | $M_P$ | 7.9° | 3.7° |
| Time to first peak | $t_P$ | 660ms | 430ms |
| Rise Time | $t_r$ | 355ms | 280ms |
| Settling Time | $t_s$ | 2780ms | 2500ms |

However, the compensator has poor robustness to HF noise - introduced by the IMU output shown in Figure 8 which is interpreted by the autopilot controller. Despite the Kalman Filter ability, additional noise is present in 'B' compensator signal seen in Figure 15 compared to 14. This was highlighted in prior research [12], and future work should address this deficiency. The compensator amplifies the noise in the system with a maximum and minimum of 0.16° & -0.23° respectively, creating chattering. Chattering is a harmful phenomenon which leads to low control accuracy and high wear of moving mechanical parts, consequently causing the elevator to fail during operation. Thus, it should be minimised or rejected.

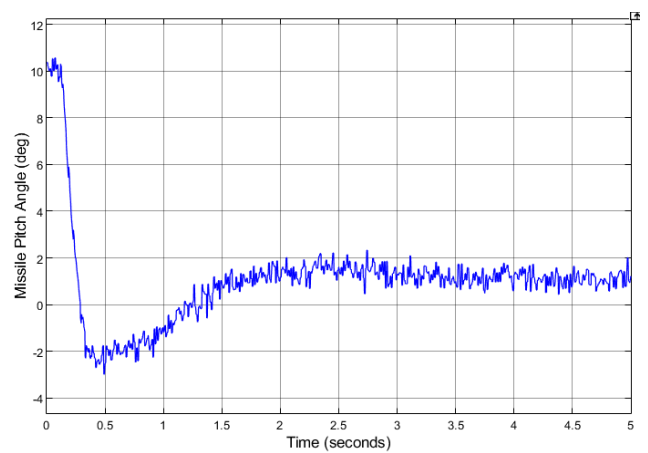

**Figure 8: Measured Pitch Angle 'B'**

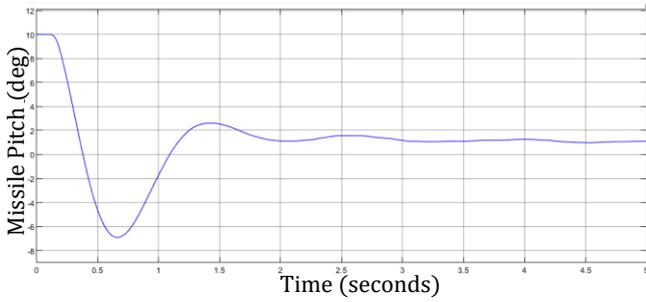

**Figure 9: Pitch Response No Compensator 'A'**

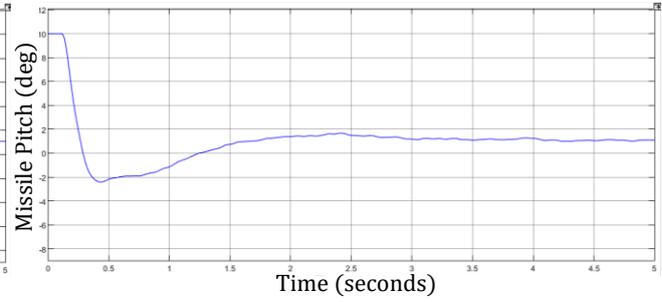

**Figure 10: Pitch Response with Compensator 'B'**

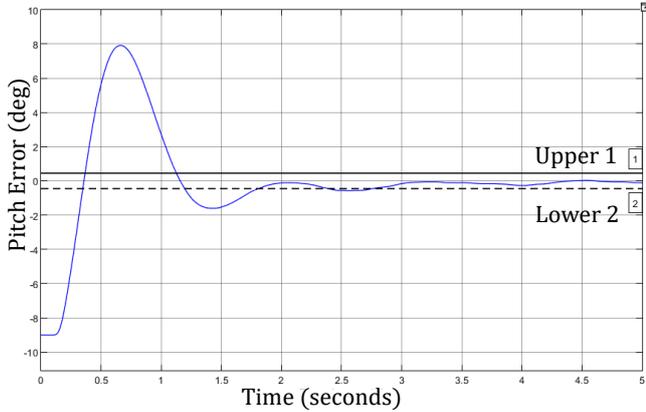

**Figure 11: Pitch Actual Error Response 'A' (Tolerance 5%)**

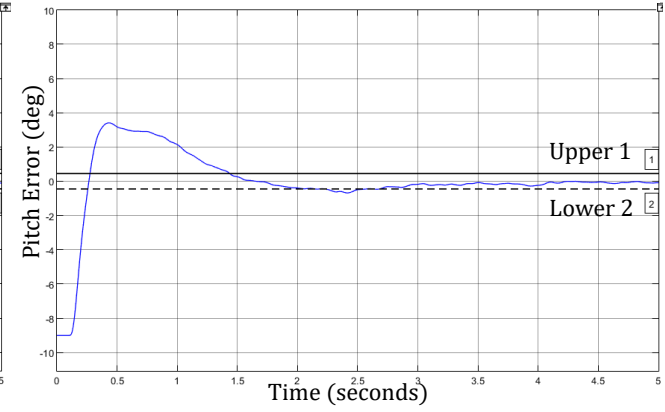

**Figure 12: Pitch Actual Error Response 'B' (Tolerance 5%)**

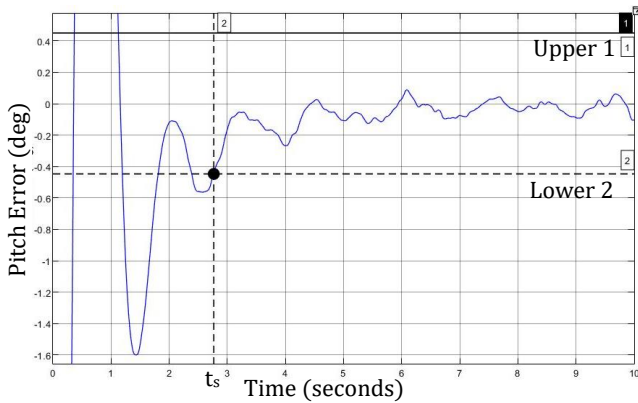

**Figure 13: Observed Amplified Noise 'A'**

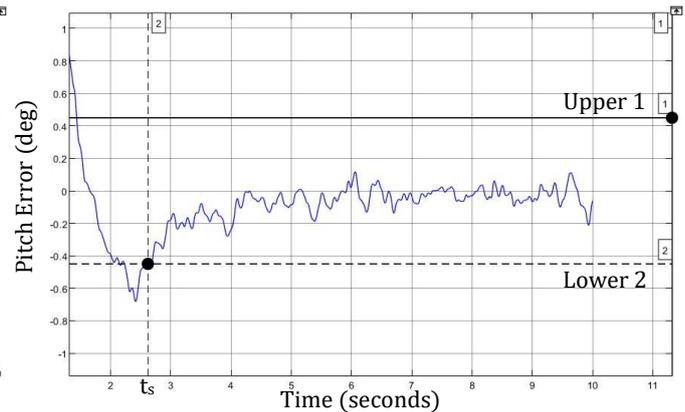

**Figure 14: Observed Amplified Noise 'B'**

The systems in Figures 9 - 14 show the stability the pitch response, as the error tends to zero well within the test time. The final system 'B' is under-damped with no additional oscillations but a 370% overshoot. The settle time $t_s$ is 10% improvement, and the rise time $t_r$ is 21% improvement on response 'A'.

E. Devaud [p2 13] states desired performances for a tail-controlled missile autopilot response to be:

(1)     $t_r \leq 350ms$
(2)     $\%M_P \leq 10\text{-}20\%$
(3)     Steady State accuracy $\leq 5\%$

Both systems 'A' & 'B' have significant overshoot that far exceeds (2), suggesting that the actuator selected is not optimum. The compensator reduced the settle time by 280ms in 'B', so the actuator delay of 100ms has been successfully negated. The steady state accuracy was factored into calculations, as response 'A' was not considered settled till 'Lower 2' was reached at 2780ms. Vice versa for response 'B' at 2500ms. Only the lead compensator system 'B', satisfies (1) and (3), with a $t_r$ =280ms and steady state response that is within 5% tolerance bands, shown in Figures 13, 15 and negates actuator delay. Only with the compensator was (1) satisfied with this designed system.

## 4.5 Hinge Moment Autopilot

The purpose of guidance is to change the missile's flight path, as defined previously with equation (6). $a_N$ is controlled via an acceleration autopilot. Several acceleration autopilots exist; however, the Hinge Moment autopilot is advantageous as it is not sensitive to dynamic pressure changes [6]. These changes would be encountered significantly when rapidly manoeuvring with 30G's within the subsonic/transonic region, and over a range of nominal operating altitudes. The autopilot should maximise $a_N \geq 30G's$ to be competitive and viable for interception.

The hinge moment transfer function for this missile is described by equation (11), requiring relevant missile data in Table 8.

$$H_0(s) = \frac{H(s)}{\delta(s)} = S_T \cdot q \cdot d_E \left( C_{M\delta}(s) + C_{M\alpha}(s) \frac{\alpha(s)}{\delta(s)} \right) \quad (11)$$

**Table 8: Missile Hinge Moment Data**

| Parameter | Note | Value | Unit |
|---|---|---|---|
| Elevator reference area | $S_T$ | 0.0865 | m$^2$ |
| Dynamic pressure @h | $q$ | 23811 | Pa |
| Velocity of missile | V | 250 | m/s |
| Length of the elevator actuator centre of pressure position to the actuator axis | $d_E$ | 69 | mm |

To ensure a smooth blending between controllers within the GNC system, a scheme called "high gain anti-windup approach" employed by (R. A. Hyde) [14] should be utilised. Research by (H. Bushek) [15] highlights the commonality of lateral and longitudinal controllers, should a 3D dynamic missile description wish to be developed.

## CONCLUSION

A subsonic missile configuration has been proposed for the Mach design point and modelled in 3D CAD using a modern software suite. Initial estimations for applicable aerodynamic properties are presented based on the closest available resources. Control research conducted within this paper shows the ability of a lead compensator to augment the pitch autopilot response ability and highlighted its HF deficiency. The autopilot satisfies 2/3 of the performance metrics, and successfully negates the actuator delay. This paper furthers the research into GNC control systems for subsonic missile interception through a 2D specific design configuration.

Future work should integrate the pitch autopilot into the hinge moment autopilot and translational model presented to create a 2D GNC system to achieve 30G's simulation interception test. This would require applying the proposed missile aerodynamic derivatives.